\def\spose#1{\hbox to 0pt{#1\hss}}

\def\multleft#1{\hbox to size{\vbox {\halign {\lft{##}\cr #1}}\hfill}\par}
\def\multright#1{\hbox to size{\vbox {\halign {\rt{##}\cr #1}}\hfill}\par}

\def\today{\ifcase\month\or January\or February\or March\or April\or May\or
      June\or July\or August\or September\or October\or November\or December\fi
      \space\number\day, \number\year}
\def\s{\hbox{\phantom{5}}}	

\def\cm{{\rm\thinspace cm}}

\def\km{{\rm\thinspace km}}
\def\kpc{{\rm\thinspace kpc}}

\def\Mpc{{\rm\thinspace Mpc}}
\def\Msun{\hbox{$\rm\thinspace M_{\odot}$}}

\def\s{{\rm\thinspace s}}

\def\kmps{\hbox{$\km\s^{-1}\,$}}

\def\pcm{\hbox{$\cm^{-3}\,$}}
\def\pcmsq{\hbox{$\cm^{-2}\,$}}

\def\psqcm{\hbox{$\cm^{-2}\,$}}

\def\kmpspMpc{\hbox{$\kmps\Mpc^{-1}$}}

\def\H2{\hbox{H$_{2}$}}
\def\Lya{Ly$\alpha$}
\def\HI{H{\footnotesize{I}}}
\documentclass[usegraphicx]{mn2e}
\newcommand{\gtsim}{\mbox{{\raisebox{-0.4ex}{$\stackrel{>}{{\scriptstyle\sim}}
$}}}}

\newcommand{\gam}{\hbox{$\Gamma$}}

\newcommand{\dex}{\hbox{\,dex}}
\newcommand{\cms}{\hbox{\,${\rm cm^{-2}}$}}
\newcommand{\civ}{\hbox{C\,{\sc iv}}}
\newcommand{\hi}{\hbox{H\,{\sc i}}}

\newcommand{\up}{\hbox{$U$}}

\newcommand{\nhi}{\hbox{$N_{HI}$}}

\newcommand{\nciv}{\hbox{$N_{CIV}$}}
\newcommand{\map}{\hbox{{\sc mappings i}c}}

\usepackage{times}
\usepackage{amssymb}
\include{defn}

\begin{document}
\hsize=6truein

\title[VLT-UVES observations of two HzRGs]{Probing the absorbing haloes around two high-redshift radio galaxies with VLT-UVES\thanks{Based on observations performed at the European Southern Observatory, Chile (Programme ID: 68.B-0086(A))}}

\author[Jarvis et al.]
{M.J.~Jarvis$^{1}$\thanks{Email: jarvis@strw.leidenuniv.nl}, R.J.~Wilman$^{1}$, H.J.A.~R\"{o}ttgering$^{1}$ and L.~Binette$^{2}$ \\
\footnotesize
$^{1}$Sterrewacht Leiden, Postbus 9513, 2300 RA Leiden, The Netherlands. \\
$^{2}$Instituto de Astronom\'ia, UNAM, Ap.70-264, 04510 M\'exico, DF, M\'exico.\\}
\maketitle

\begin{abstract}
We present VLT-UVES echelle spectroscopy of the \HI~ and
\civ~absorption in the spatially-extended \Lya~emission around two
high-redshift radio galaxies 0200+015 (z=2.23) and 0943-242
(z=2.92). 

The absorbers in 0943-242 exhibit little additional structure compared
with previous low-resolution spectroscopy and the main absorber is
still consistent with H{\footnotesize{I}}~column density of $\sim
10^{19}$\pcmsq. This is consistent with a picture in which the
absorbing gas has low density and low metallicity and is distributed
in a smooth absorbing shell located beyond the emission-line gas.
However, the main absorbers in 0200+015 are very different. The
previous single absorber fit of \HI~ column density $\simeq
10^{19}$\pcmsq, now splits into two $\sim 4 \times
10^{14}$\pcmsq~absorbers which extend more than 15\kpc~to obscure
additional \Lya~emission coincident with a radio lobe in these
high-resolution observations. Although consistent with the shell-like
distribution for the absorption systems, 0200+015 requires a much
higher metal enrichment than 0943-242. The metallicity, inferred from
the \civ~ absorption, is considerably lower in 0943-242 than in
0200+015.  We explain these differences with an evolutionary scenario
based on the size of the radio source.  In both sources the \HI~
absorption gas originates from either a gas-rich merger or pristine
cluster gas which cools and collapses towards the centre of the dark
matter halo.  The higher metallicity in the larger radio source
(0200+015) may be a result of a starburst driven superwind (concurrent
with the triggering of the radio emission) which has engulfed the
outer halo in this older source.

We also find a significant blue asymmetry in the HeII$\lambda 1640$ emission
line, suggesting that the line emitting gas is outflowing from the
central regions. Dust obscuration toward the central engine,
presumably due to the dusty torus invoked in Unified Scheme, prevents
us from seeing outflow away from our line-of-sight.

\end{abstract}


\section{INTRODUCTION}
The existence of powerful radio galaxies at high redshift ($z>2$)
 demonstrates that a population of supermassive black holes ($>
 10^{9}~\Msun$; Dunlop et al. 2002) was in place just a few billion
 years after the Big Bang. Like their counterparts at low redshift,
 the high-redshift radio galaxies (HzRGs) appear to reside in the most
 massive elliptical galaxies at their epoch (e.g. Jarvis et
 al.~2001b), in line with the established correlation between black
 hole and spheroid mass (Magorrian et al.~1998). Whilst the discovery
 of radio quiet galaxies at high redshift via Lyman drop-out
 techniques means that they are no longer our only probe of high
 redshift galaxy formation, the HzRGs still provide the most important
 insights into the early formation of the most massive bound
 structures. Indeed, the depth of the gravitational potential wells in
 which they form renders them prime targets for searching for
 high-redshift protoclusters, as confirmed by the recent discoveries
 of over-densities of \Lya~ emitters around such objects (see
 e.g. Kurk et al.~2000; Pentericci et al.~2000; Venemans et al.~2002).

One of the most prominent characteristics of HzRGs are their extended
 emission line regions (EELRs), which are luminous ($>10^{37}$~W in
 \Lya), tens to several hundred kpc in size (often aligned with the
 radio axis) and kinematically active (FWHM~$\gtrsim
 1000$\kmps). Their dominant ionisation mechanism seems to change from
 shocks to photoionisation by the central engine as the radio source
 expands beyond the confines of the host galaxy (Best et al.~2000; De
 Breuck et al.~2000; Jarvis et al. 2001a). Likewise the kinematics of
 the gas may reflect the competing influences of gravity, and energy
 input via shocks from the radio source and starburst-driven
 superwinds. However, there are many unresolved issues which remain
 concerning the structure, origin and fate of the emission line gas,
 which may hold important clues to many aspects of massive galaxy
 formation. Did the gas originate in a cooling flow or a merger, or
 was it expelled from the central galaxy during a violent starburst?
 Is it in the form of cloudlets, filaments or expanding shells of
 material, and what is their composition?
What is the ultimate fate of the gas, will it
 form stars or escape from the galaxy to enrich the surrounding
 intracluster and intergalactic media?

A new perspective on many of these issues was opened up with the
discovery by R\"{o}ttgering et al.~(1995) and van Ojik et al.~(1997; hereafter vO97)
that most of the smaller high redshift radio sources (those with projected linear size $D <50$\kpc) exhibit spatially resolved \HI~ absorption in their
\Lya~ emission-line profiles, with column densities in the range
$10^{18}$--$10^{19.5}$\pcmsq. Binette et al.~(2000; hereafter B00) found, for the $z=2.92$ radio galaxy 0943$-$242, \civ$\lambda\lambda$1548,1551 absorption
lines superimposed on the \civ~ emission, at the same redshift as the
main \Lya~ absorption system. They could not reconcile the
observed \civ/\Lya~ emission- and absorption-line ratios with a model in
which the absorption- and emission-line gas are co-spatial. Instead
they proposed that the absorbing gas is of lower metallicity ($Z \sim
0.01 Z_{\rm{\sun}}$) and located further away from the host galaxy
than the emission line gas, beyond the high pressure radio source
cocoon. This material, they claimed, is thus a relic reservoir of low
metallicity, low density ($\sim 10^{-2.5}$\pcm) gas, similar to that
from which the parent galaxy may have formed.  This shell-like distribution of
the absorbing gas was also invoked by Jarvis et al. (2001a) and De Breuck et al. (2000) to explain
various correlations found in a complete sample of high-redshift ($z >
2$) radio galaxies. 

In this paper we probe the H{\footnotesize{I}} and \civ~ absorbers in greater detail using the UVES
spectrograph on the VLT to obtain echelle spectra of two high-redshift
radio galaxies, the aforementioned 0943-242 and 0200+015 at
z=2.23. 
The new observations offer a factor of 10 improvement in spectral
resolution over previous investigations and thus have the potential
(i) to reveal whether the main absorbers are genuinely in the form of
a single component with ordered global kinematics, or whether they
fragment into a number of weaker absorbers; (ii) to probe the
emission-line kinematics with high velocity resolution.

In section~\ref{sec:obs} we describe the observations and data
reduction of our VLT-UVES spectroscopic observations and in
section~\ref{sec:analysis} we briefly discuss the emission- and
absorption-line fitting procedure. In section~\ref{sec:0943} we use
our new high-resolution spectra to investigate the absorption halo of
the $z = 2.92$ radio galaxy 0943-242. Section~\ref{sec:0200} is a
detailed discussion of the absorbing halo around the $z = 2.23$ radio
galaxy 0200+015 and we also investigate the
kinematics of the narrow-emission lines via the unabsorbed
HeII emission line and briefly discuss the effect of emission-line
asymmetry on our absorption line fitting. The origin and fate of the
absorbing halos is discussed in section~\ref{sec:discussion} and we
provide a summary of our conclusions in section~\ref{sec:conc}.

All physical distances are calculated assuming $H_{\rm{0}}=70$\kmpspMpc,
$\Omega_{\rm{M}}=0.3$ and $\Omega_{\rm{\Lambda}}=0.7$.

\section{OBSERVATIONS AND DATA REDUCTION}\label{sec:obs}
The observations were performed with the UVES echelle spectrograph on
VLT UT2 at the European Southern Observatory on the night of 2001
December 8-9. For 0200+015, dichroic 1 was used with cross-disperser 1
in the blue arm (central wavelength of 3950\AA, for \Lya), and
cross-disperser 3 in the red arm (central wavelength of 5800\AA, for
\civ$\lambda\lambda$1548,1551 and HeII$\lambda$1640); three separate
exposures of 1 hour were taken. For the 0943-242, just the red arm was
used, with cross-disperser 3 set to a central wavelength of 5200\AA,
sufficient to cover both \Lya~and
\civ$\lambda\lambda$1548,1551. Separate exposures totalling 3.4 hours
were acquired. For all observations on-chip binning of $2 \times
3$~(spatial $\times$ spectral) was used, resulting in a pixel size of 0.5~arcsec in the
red arm, and 0.36~arcsec in the blue. The resulting pixel scales in
the dispersion direction for the various set-ups were in the range
0.05-0.06\AA. Together with a slit-width of 1.2~arcsec, the resulting
spectral resolutions were 25000-40000. The seeing was steady
throughout the night at $\sim 0.8$~arcsec, and the sky dark and
photometric. The slit position angles were chosen to align with the
radio axes and previous observations, namely 155 degrees for 0200+015
and 74 degrees east of north for 0200+015 and 0943-242 respectively.

Offline reduction was performed with IRAF, following the procedures
for echelle data described by Churchill~(1995). For each target, the
individual exposures were median combined to remove cosmic rays,
bias-subtracted and flat-fielded. The order-definition frames were
used to determine the locations of the orders, which were then
extracted and wavelength calibrated using a ThAr arc and combined. The
blaze function of the grating was removed using observations of the
G-type spectrophotometric standard LTT~1020. Spectra for individual
spatial regions along the slit were extracted, rebinned by a factor of
2 in dispersion to improve the signal-to-noise ratio.

\section{ANALYSIS}\label{sec:analysis}
The extracted 1-d spectra were fitted with a series of Voigt
absorption line profiles superimposed upon a Gaussian emission
envelope. We choose an underlying Gaussian emission-line profile as
studies of narrow-line profiles in Seyfert-2 galaxies show that this
is preferred over a Lorentzian profile in the majority of cases
(e.g. V\'eron-Cetty, V\'eron \& Gon\c{c}alves 2001), but see section~\ref{sec:emlines} for further discussion of this point. The model profiles were convolved with the instrumental
function prior to fitting (as described by R\"{o}ttgering et al.~1995 and B00). There is no detected continuum underlying the emission lines.

\section{0943-242: Conclusive evidence for a distinct shell of absorption-line gas?}\label{sec:0943}

\begin{figure}
\includegraphics[width=0.48\textwidth,angle=0]{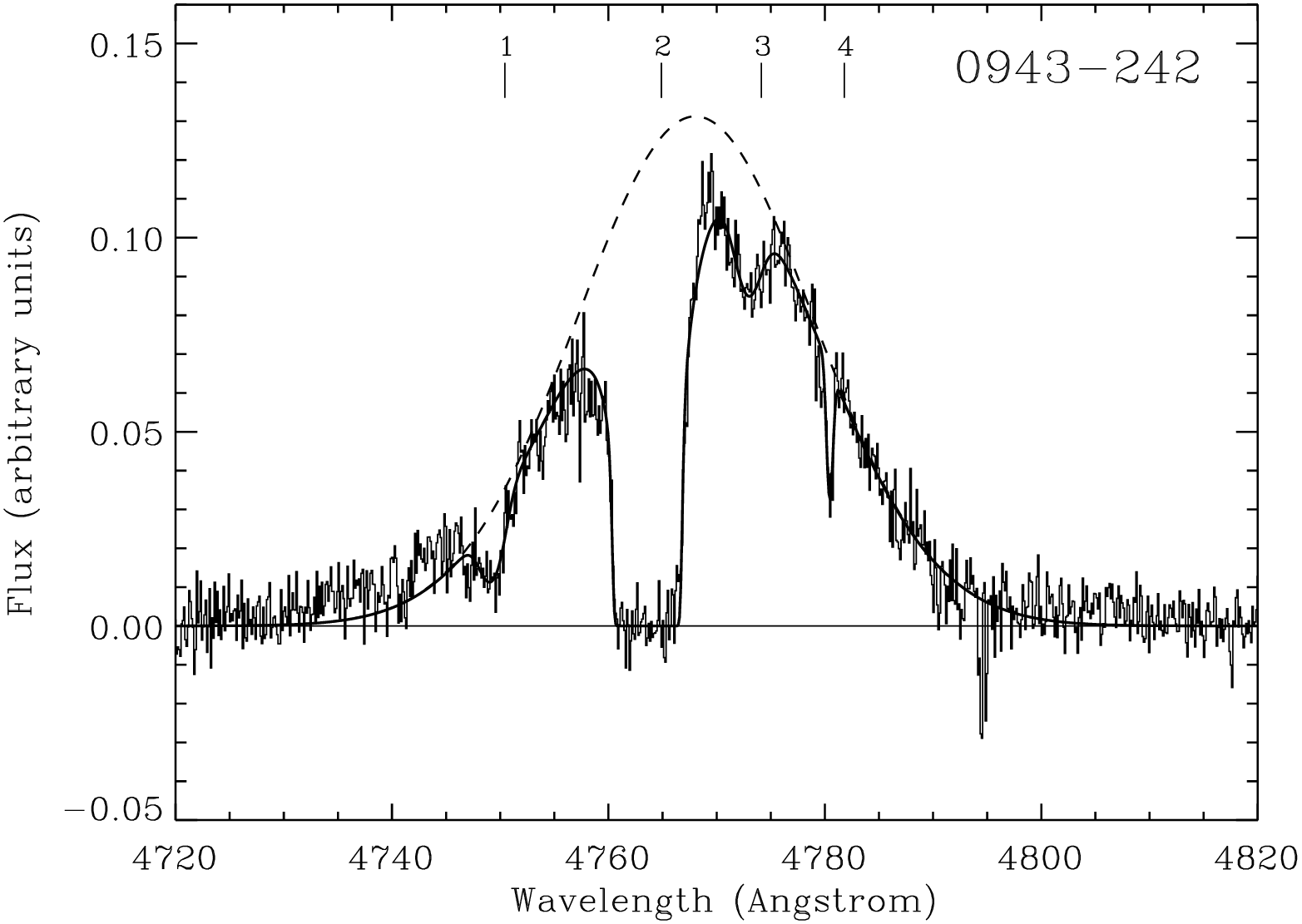}

\includegraphics[width=0.48\textwidth,angle=0]{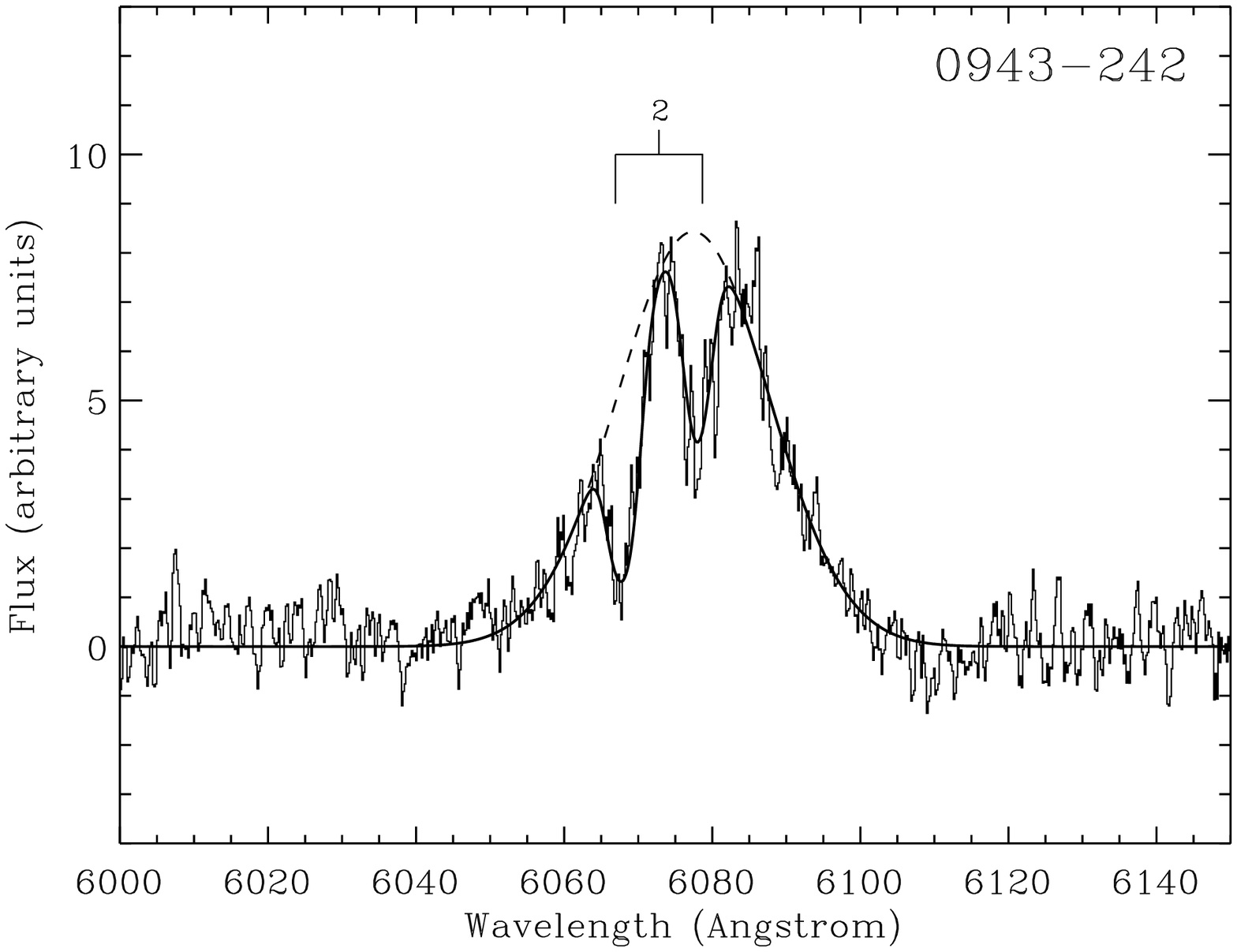}
{\caption{\label{fig:Lya0943}\normalsize ({\it top}) The \Lya~profile of 0943-242, with the absorption model overlaid. ({\it bottom}) The CIV profile of 0943-242, with the absorption model overlaid.  }}
\end{figure}

Fig.~\ref{fig:Lya0943} shows the 1-d spectrum of the \Lya~ and \civ~emission lines in 0943-242,
extracted over the central 5~arcsec of the slit which covers all of
the emission. The fit, with a series of Voigt profile absorbers
superimposed upon a Gaussian emission envelope is shown, and the
parameters tabulated in Table~\ref{tab:Lya0943}. The errors were
calculated by assuming a $\chi^{2}$ distribution of $\Delta \chi^{2}$
($\equiv \chi^{2} - \chi^{2}_{\rm min}$) (Lampton, Margon \& Bowyer
1976). The error on each parameter was calculated by setting $\Delta
\chi^{2} = 1$ and allowing the other parameters to float. The errors
quoted are $1\sigma$.

The most notable
feature of this spectrum is that the main absorber
remains as a single system of column density $\sim 10^{19}$\psqcm, and
is completely black at its base with no evidence for substructure. No
new absorption systems are identified in addition to those found by
R\"{o}ttgering et al.~(1995) at a factor of $\sim 10$ lower
resolution.

This provides strong evidence that these absorption systems
are physically distinct from the extended narrow-emission-line
region. If the absorption was caused by gas which is mixed with the
emission-line clouds, as postulated by vO97 then we
would expect a series of narrow-absorption troughs at various
wavelengths across the \Lya~ emission profile. The fact that we see one
main absorber, with no detectable substructure, shifted blueward of
the emission-line peak provides compelling evidence that the
absorption gas encompasses the whole of the emission-line regions with
a covering factor of unity along our line-of-sight. This is in line with the argument proposed
by B00 who fitted the emission- and absorption-line
profile of the \civ$\lambda\lambda$1548,1551 doublet in 0943-242 to
determine the metallicity of the absorption systems compared to the
emission-line gas. Their results show that the emission and absorption
line ratios of \civ~ and \Lya~ are incompatible with photoionisation or
collisional ionisation of cloudlets with uniform properties and the
possibility that the absorption and emission phases are co-spatial is
rejected. A different model was preferred in which the absorption
gas has low metallicity and is located further away from the host
galaxy than the emission line gas. 

\begin{table}
\begin{tabular}{|ccll|} \hline
Absorber  & $z$ & $b$ & log $N(\rm{HI})$ \\
 & & (\kmps) & (\psqcm) \\  \hline
1  & 2.9066 $\pm$ 0.0062 & 88 $\pm$ 45  & 14.02 $\pm$ 0.30 \\
2  & 2.9185 $\pm$ 0.0001 & 58 $\pm$ 3  & 19.08 $\pm$ 0.06 \\
3  & 2.9261 $\pm$ 0.0005 & 109 $\pm$ 35 & 13.55 $\pm$ 0.16 \\
4  & 2.9324 $\pm$ 0.0001 & 23 $\pm$ 17 & 13.35 $\pm$ 0.30 \\ 
\hline
 &  &  & log $N(\rm{CIV})$ \\
& & & (\psqcm) \\
    \hline
2  & 2.9192 $\pm$ 0.0001 & 119 $\pm$ 2 & 14.58 $\pm$ $\pm$ 0.04 \\
\hline
\end{tabular}
\caption{\label{tab:Lya0943} Parameters of the Voigt absorption profile fits for the main component of 0943-242. The final row corresponds to the fit of one absorber to the CIV profile.}
\end{table}

\section{0200+015: A detailed study}\label{sec:0200}

In this section we provide an in depth study of the absorption systems around the $z = 2.23$ radio galaxy 0200+015. This object was taken from the original sample of vO97 and was
chosen on account of its interesting absorption-line profile in the
$\sim 2$\AA\, resolution spectrum of vO97. A further advantage of
using this object for our study is that its redshift permits further
observations of all of the principal optical emission lines in the near-infrared transmission windows arising
from AGN phenomena which will enable us to determine line ratios and
investigate the ionisation mechanisms at work as a function of radial
distance from the central source.

\begin{figure}
\includegraphics[width=0.48\textwidth,angle=0]{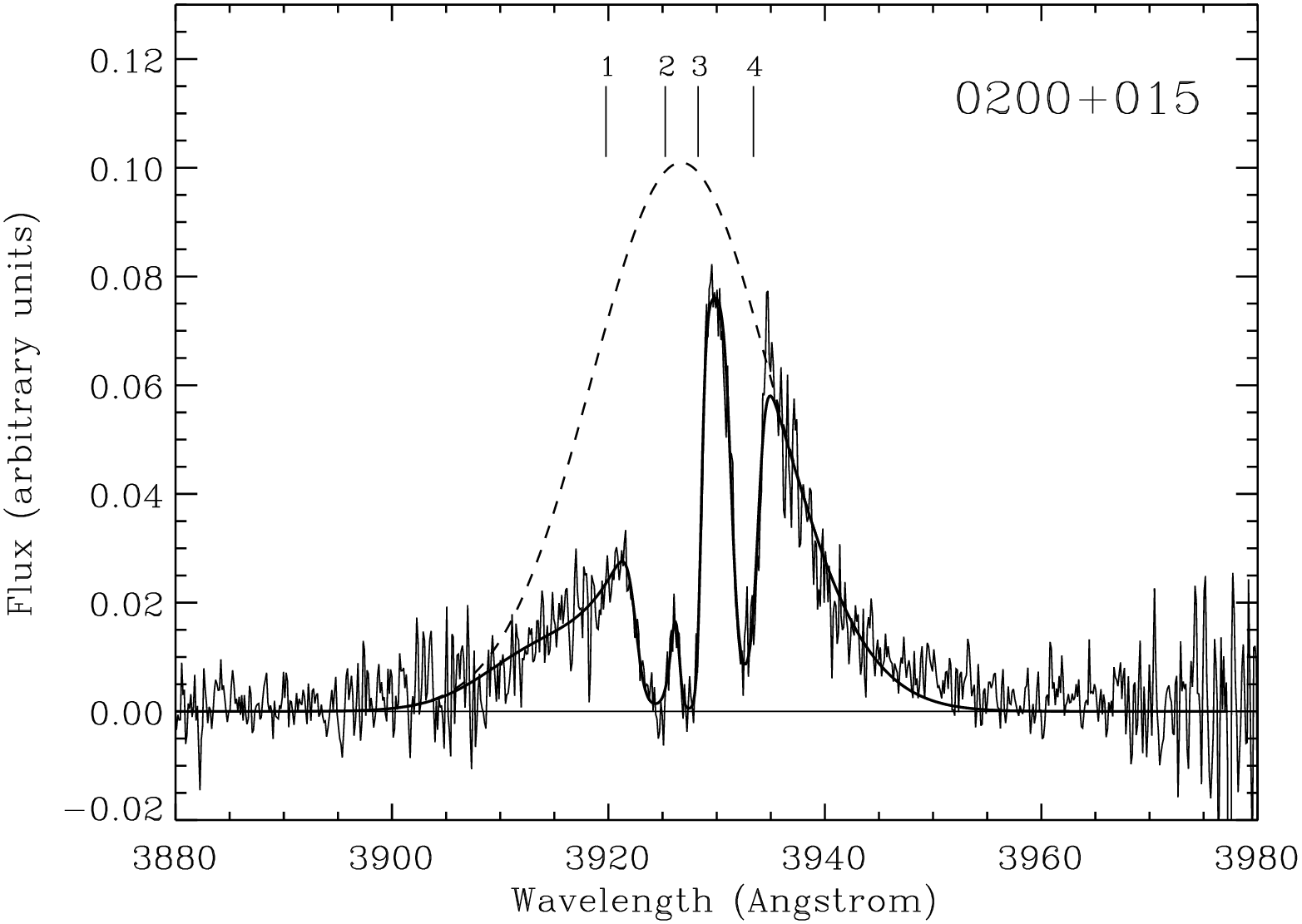}
\includegraphics[width=0.48\textwidth,angle=0]{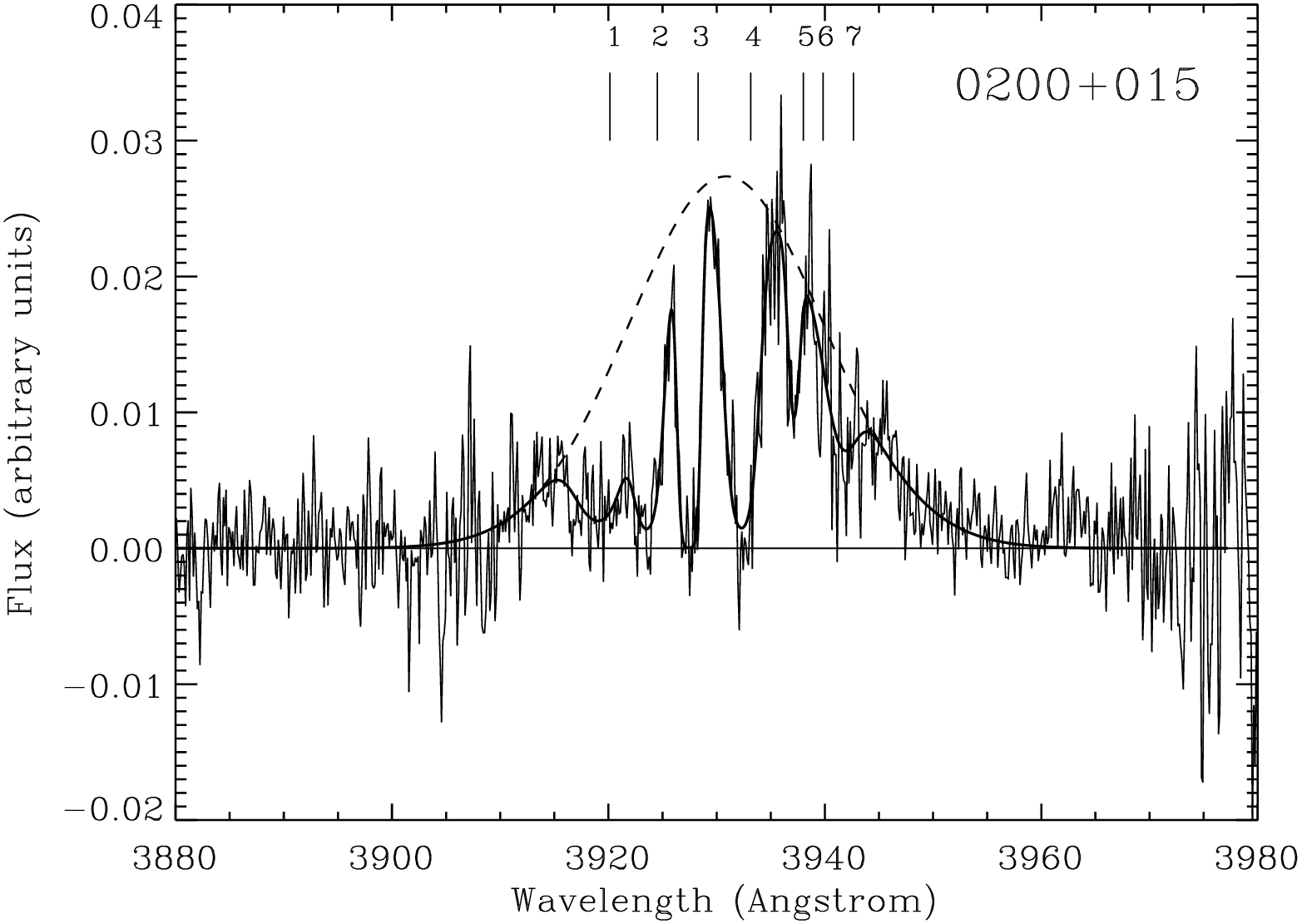}
\caption{\label{fig:Lya0200lobe}\normalsize (a) ({\it top}) The \Lya~profile of the central 2.5~arcsec of 0200+015, with the Gaussian emission envelope and the absorption model overlaid. (b) ({\it bottom}) As in (a) but for the secondary (offset) component of 0200+015, with the absorption model overlaid.}
\end{figure}

Examination of the \Lya~region of the CCD frame of this object shows
that the emission deviates from the centre of the order (as defined by
the order definition frame) as a function of wavelength. This reflects
spatial extent in the emitting and absorbing gas. Spectra were
extracted for 2 spatial regions: the main body of the galaxy between
$\pm1.25$~arcsec of the centroid, and a secondary component lying
between 1.25--2.5~arcsec to the north-west (1 arcsec translates into
8\kpc~at this redshift), coincident with one of the two radio lobes
(see the radio continuum image in Carilli et al.~1997). Spectra with
overlaid absorption models for these two components are shown in
Fig.~\ref{fig:Lya0200lobe}, and the parameters tabulated in Tables~\ref{tab:Lya0200main} and \ref{tab:Lya0200lobe}.

\subsection{The \Lya~ emission from the nucleus}

\begin{table}
\caption{\label{tab:Lya0200main}Parameters of the Voigt absorption profile fits for the main component of 0200+015}
\begin{tabular}{|ccll|} \hline
Absorber & $z$ & $b$ & log $N(\rm{HI})$ \\
 & & (\kmps) & (\psqcm) \\  \hline
1 & 2.2239 $\pm$ 0.0007 & 612 $\pm$ 70 & 14.96 $\pm$ 0.06 \\
2 & 2.2282 $\pm$ 0.0002 & 116 $\pm$ 27 & 14.73 $\pm$ 0.14 \\
3 & 2.2307 $\pm$ 0.0001 & 61 $\pm$ 8 & 14.58 $\pm$ 0.16 \\
4 & 2.2349 $\pm$ 0.0001 & 86 $\pm$ 11 & 14.38 $\pm$ 0.10  \\ \hline
\end{tabular}
\end{table}

The core \Lya~ emission-line profile (i.e. the emission associated with
the centre of the radio galaxy host) has a redshift of $z = 2.2292$ and full width half maximum FWHM = 1490~\kmps. The emission line exhibits a number of absorbers
within $320$~\kmps~ of the redshift of the peak of the Ly$\alpha$ emission, consistent with the
lower resolution spectrum of vO97. However, at this
higher resolution we do find that the main absorber, with a measured
column density of $N_{\rm HI} = 10^{19.1}$~cm$^{-2}$  from vO97, is actually comprised of two
absorption troughs with column densities of $N_{\rm HI} =
10^{14.73}$~cm$^{-2}$ and $N_{\rm HI} = 10^{14.58}$~cm$^{-2}$ with 
velocity widths of 116\kmps~and 61\kmps~ and
redshifts of $z = 2.2282$ and $z= 2.2307$ respectively. 

There is also a $N_{\rm HI} = 10^{15}$~cm$^{-2}$
absorber blueward of these lines if our Gaussian emission profile is
accurate. This absorption trough has a much higher Doppler parameter
of $b = 612$\kmps, although the sensitivity of this absorption profile on the fit to the emission line is significant and we explore this further in section~\ref{sec:emlines}.

\subsection{Emission-line asymmetry and the evidence for outflow}\label{sec:emlines}
\begin{figure}
\includegraphics[width=0.48\textwidth,angle=0]{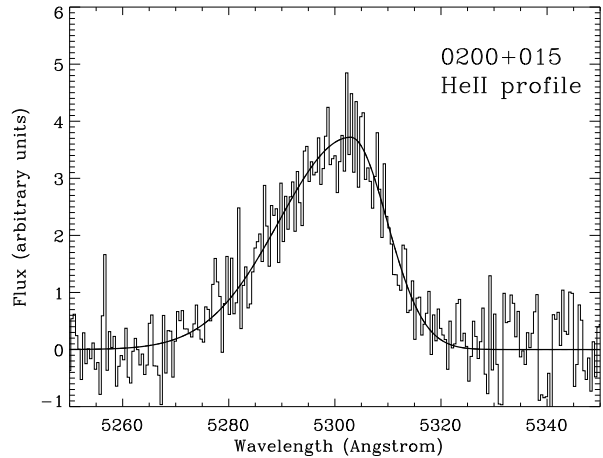}
\caption{\label{fig:he2}\normalsize The HeII profile of the central 2.5~arcsec of 0200+015.}
\end{figure}

One of the very noticeable aspects of the spectroscopic data on these
two high-redshift radio galaxies is that in the HeII emission line, in
which associated absorption does not affect the profile of the
emission line, the blue side of the emission line has a longer tale
than the redward side (Fig.~\ref{fig:he2}). Using the profile analysis
originally defined by Heckman et al. (1981) we assign a relative
asymmetry, AI, to the emission at varying fractions of the total
intensity where AI $= 2[\Lambda_{\rm c} -
\lambda_{\rm c}(i)]/\omega(i)$. Here $\lambda_{\rm c}(i)$ and
$\omega(i)$ are the line centre and width at relative intensity $i$,
respectively, and $\Lambda_{\rm c}$ is the wavelength of the peak of
the profile. The results of the profile analysis are presented in
Table~\ref{tab:he2_ass}.  Using the FWHM of the asymmetric profile to
determine the velocity dispersion of this line we find $v \approx
1920$\kmps.

This blue-tail line asymmetry has been found previously for the
narrow-line regions in many Seyfert galaxies (e.g. De Robertis \& Shaw
1990) and is probably associated with an inflow or outflow of the gas
into or from the central regions. Evidence from
narrow-line Seyfert galaxies points towards the view that the
asymmetry is caused by outflowing gas propagating towards us along our
line-of-sight. The counter flow, which would produce a red-tail
asymmetry, is obscured by dust residing within the radius of the
narrow-emission-line region. This is in line with the dusty torus invoked in orientation based Unified
Schemes.
The alternative view, that of inflowing gas, requires the dust to be
mixed with the emission-line clouds. This would mean that the blue
tail emanates from the far side of the galaxy. In this case the illuminated
clouds on the near side along our line of sight suffer higher
extinction than the illuminated clouds on the far side (e.g. De
Robertis \& Shaw 1990).
However, ISO observations of NGC~4151 show that this explanation is
not sufficient to explain the emission-line characteristic in the
infrared where the emission lines are much less susceptible to
obscuration.  Sturm et al. (1999) propose a central, optically thick
obscuring screen close to the nucleus with outflowing emission-line
regions, most plausibly associated with a nuclear outflow or a
starburst.

\begin{table}
\begin{center}
\caption{\label{tab:refit} Asymmetry of the HeII line in 0200+015}
\begin{tabular}{|clll|} \hline
per cent of peak ({\it i}) & $\lambda_{\rm c}$ / \AA & $\omega$ / \AA & AI \\
\hline
10 & 5296.4 & 43.6 & 29.4 \\
20 & 5296.9 & 36.1 & 32.4 \\
33 & 5298.1 & 29.9 & 31.8 \\
50 & 5298.9 & 23.0 & 33.9 \\
\hline
\end{tabular}
\end{center}
\end{table}

\begin{figure}
\includegraphics[width=0.48\textwidth,angle=0]{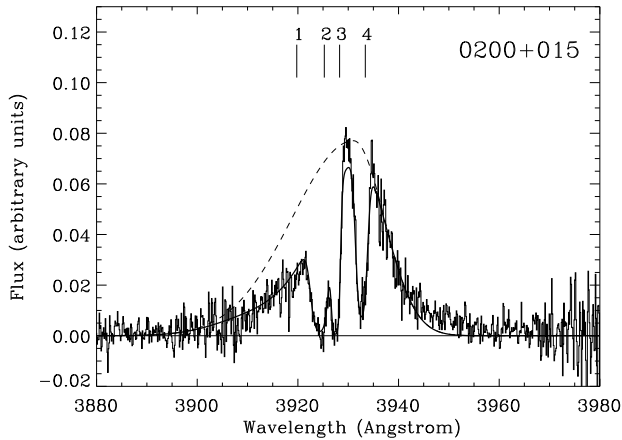}
\caption{\label{fig:0200Lyaasymm}\normalsize The Ly$\alpha$ profile of the central 2.5~arcsec of 0200+015 fitted with an asymmetric Gaussian with the same asymmetry as the HeII emission line .}
\end{figure}

Obviously, this particular characteristic of
the emission line region needs exploring further with high-resolution
spectroscopy of emission lines with a range in ionisation parameters
to determine whether the velocity of the material is dependent on the
density of the gas and its distance from the central ionizing
source. However, for the purposes of this paper where we are
predominantly concerned with the absorbing shells, it highlights the
fact that assuming a simple Gaussian profile for the narrow emission
lines may be too simple. We have therefore refitted the \Lya~ and
\civ~ emission lines for our two objects to determine whether this
amount of asymmetry may significantly affect our fitted parameters to
the absorption troughs in any way.

We refitted our emission-line profiles with two single tailed
Gaussians with common peak intensities and a fixed ratio of FWHM
derived from the asymmetry in the HeII line as a simple approximation
to the asymmetry profile. Although we note that the different
ionisation potential for \Lya~ and \civ~ compared to HeII may affect
the asymmetric profile, we do not expect this to be a significant
difference given the uncertainties in our model fitting. With this
fixed asymmetry the resulting fits to the absorption troughs in both
\Lya~ and \civ~ are not significantly altered, these fitted parameters
are listed in Table~\ref{tab:refit}

We therefore conclude that although there is evidence for line
asymmetry in the narrow-emission-lines in HzRGs this does not alter
the fits to the absorption troughs significantly and the original
results are still valid. This is quite surprising given the
uncertainties arising from the emission-line profile. However, we note
that it is very difficult to change the actual absorption columns
given any reasonable Ly$\alpha$ emission-line profile. This is
particularly true where the emission is at a low level, i.e. around
absorber \#1 in Fig.~\ref{fig:Lya0200lobe} and
Fig.~\ref{fig:0200Lyaasymm}. Whereas the largest errors occur near the
peak of the emission where the absorbing columns (absorbers \#2 and
\#3) have larger associated errors. Furthermore, the fits to absorbers
\#2 and \#3 are more tightly constrained in $b$ because the higher
degree of absorption at lower $b$ means that there is little room to
manouvre in the absorption line fitting. The combination of these
effects results in little variation in the absorptions line column
densities. However, we note that more extreme Ly$\alpha$ emission-line
profiles will result in larger errors on the absorption-line fits in
the wings of the emission line.

\begin{table}
\begin{center}
\caption{\label{tab:he2_ass} Parameters of the Voigt absorption profile fits for 0200+015, assuming asymmetric emission-line profile using the asymmetry parameters from the HeII emission line. The top panel is for the four absorbers in the Ly$\alpha$ emission line from the nucleus , the lower panel represents the single absorber in the CIV profile.}
\begin{tabular}{|lllll|} \hline
 Absorber & Line & $z$ & $b$ & log $N(\rm{HI})$ \\ & & & (\kmps) &
(\pcmsq) \\
\hline
 1 &\Lya & 2.2206 $\pm$ 0.0012 &  874 $\pm$ 130 & 14.93 $\pm$ 0.07 \\
 2 &\Lya & 2.2282 $\pm$ 0.0003 &  120 $\pm$ 25 & 14.71 $\pm$ 0.12 \\
 3 &\Lya & 2.2307 $\pm$ 0.0001 &  55 $\pm$ 7 &  14.52 $\pm$ 0.17 \\
 4 &\Lya & 2.2349 $\pm$ 0.0001 &  83 $\pm$ 11 &  14.39 $\pm$ 0.09 \\
\hline
&  & & & log $N(\rm{CIV})$ \\
&  & & &   (\pcmsq) \\
\hline
 2 & \civ & 2.2285 $\pm$ $<$0.0001 & 59 $\pm$ 2 & 14.64 $\pm$ 0.01 \\
\hline
\end{tabular}
\end{center}
\end{table}

\subsection{The \Lya~ emission associated with the radio lobe}

We also observe relatively bright \Lya~ emission at the position of
one of the radio lobes associated with 0200+015. This emission lies
$\approx 2.5$~arcsec ($\approx 20$~kpc) away from the nuclear
emission.  The best-fit Gaussian profile to this emission line shows
that the emission has a slightly higher redshift than that found
towards the central radio nucleus ($\delta v \sim 150$~\kmps) and has a FWHM = 1580~\kmps.

This emission is presumably due to the passage of the radio emitting
plasma disrupting the gas at these radii. Invoking the scenario set
out by Jarvis et al. (2001a) based on the model of Bremer, Fabian \&
Crawford (1997), in which the very extended narrow-emission-line
regions are compressed and shredded by the radio jets, then this line
emission may have been associated with the innermost low-density
absorbing shell, which has condensed enough to be seen in
emission. This would also explain the slight velocity shift of this
emission line if the radio jet, which has essentially boosted this
emission, is propagating away from our line-of-sight.  However,
regardless of the process by which the
\Lya~ photons are emitted, the surrounding absorbers are distributed in
the same way as we see in the nuclear \Lya~ emission. This is probably
the most compelling evidence that the absorbing gas is distributed in
a shell encompassing the whole of the radio galaxy with a covering factor of $\approx 1$, in line with our
conclusions for 0943-242. The velocity difference of the absorption
systems detected in the \Lya~ emission associated with the nuclear
emission and that at the position of the lobe is $\sim 30$~\kmps, thus
we can be quite confident that the absorption systems seen in the lobe emission
are associated with the absorption systems associated with the nuclear
emission.

We also find more absorbers in this extended emission line in addition
to the four main absorbers found in the nuclear \Lya~ emission
line. These additional absorbers are on the red side of the Ly$\alpha$ 
emission line and have significantly lower column densities. The largest
velocity offset from the Ly$\alpha$ emission-line redshift for these redshifted
absorption systems is $\sim 600$~km~s$^{-1}$. The lower measured
column densities of the absorption gas, implies that the covering
factor is lower along the line-of-sight to the emission associated with the lobe. 

If we assume that the emission from the lobe, which is
redshifted with respect to the Ly$\alpha$ emission-line  redshift at the nucleus, is on
the far side and is propagating away from us, then any emission from
this position would have to pass through both a lower column density
redshifted \HI~ component associated with the far side and the bulk of
the gas on the near side which is blueshifted. 
This would explain the
higher column density absorbing gas on the blue wing of the emission
line. 

\begin{table}
\caption{\label{tab:Lya0200lobe} Parameters of the Voigt absorption profile fits for the secondary component of 0200+015}
\begin{tabular}{|clll|} \hline
Absorber & $z$ & $b$ & log $N(\rm{HI})$ \\
 & & (\kmps) & (\psqcm) \\  \hline
1  & 2.2241 $\pm$ 0.0010 & 216 $\pm$ 110 & 14.68 $\pm$ 0.26 \\
2  & 2.2276 $\pm$ 0.0002 & 106 $\pm$ 29 & 14.52 $\pm$ 0.22 \\
3  & 2.2307 $\pm$ 0.0001 & 61 $\pm$ 11  & 14.70 $\pm$ 0.24 \\
4  & 2.2347 $\pm$ 0.0002 & 111 $\pm$ 21 & 14.63 $\pm$ 0.16 \\
5  & 2.2387 $\pm$ 0.0003 & 44 $\pm$ 37 & 13.69 $\pm$ 0.35 \\
6  & 2.2402 $\pm$ 0.0008 & 24 $\pm$ 22  & 13.23 $\pm$ 0.65 \\
7  & 2.2423 $\pm$ 0.0018 & 87 $\pm$ 59  & 13.95 $\pm$ 0.58 \\ 
\hline
\end{tabular}
\end{table}

\subsection{The CIV emission from the nucleus}

\begin{figure}
\includegraphics[width=0.48\textwidth,angle=0]{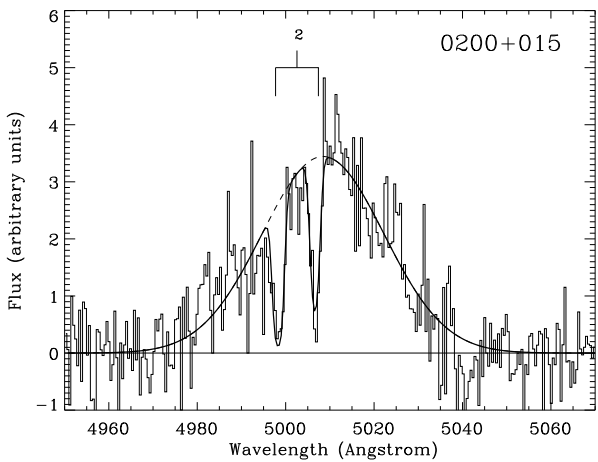}
\caption{\label{fig:civ0200}\normalsize The CIV profile of 0200+015, with the vertical ticks indicating the location of the absorption doublet.}
\end{figure}

\begin{table}
\caption{\label{tab:civ0200} Parameters of the Voigt absorption profile fits for the CIV component of 0200+015.}
\begin{tabular}{|ccll|} \hline
Absorber & $z$ & $b$ & log $N(\rm{CIV})$ \\
 & & (\kmps) & (\psqcm) \\  \hline
2 & 2.2285 $\pm$ $< 0.0001$ & 69 $\pm$ 1 & 14.69 $\pm$ 0.02 \\
\hline
\end{tabular}
\end{table}

In the red arm of the spectrum, emission lines of the
\civ$\lambda\lambda1548$\AA,$1551$\AA~ doublet and HeII$\lambda1640$\AA~ are
present. The former is shown in Fig.~\ref{fig:civ0200}, extracted
across the entire galaxy because no spatial structure was
evident\footnote{Using the \Lya:~\civ~ emission-line flux ratio at the
nucleus, our observations are not sensitive enough to detect the
expected flux in \civ~ at the position of the radio lobe.} and rebinned in
wavelength by a factor of 10 to improve the signal to noise ratio. From our line fitting we find that the centre of the emission line is redshifted with respect to the nuclear Ly$\alpha$~emission line by $\sim 140$~\kmps, however this may be an artifact of trying to fit a symmetric line profile (cf. section~\ref{sec:emlines}) with a FWHM = 1940~\kmps. 

A clear absorption doublet can be identified, corresponding to
\civ$\lambda\lambda1548$\AA,$1551$\AA~ at $z \simeq 2.2285$, i.e. roughly
corresponding to absorption system \#2 in \Lya~ (see
Tables~\ref{tab:Lya0200main} and ~\ref{tab:Lya0200lobe}). \civ~
absorption due to \Lya~absorber \#3 would appear at $\simeq
5001$\AA,$5010$\AA~ but no obvious signature is present. \Lya~
absorber \#4 would appear at $\simeq 5007$\AA,$5017$\AA~ respectively,
thus the 1548\AA~ trough would overlap on the 1551\AA~ trough of
absorber \#2. However, the lack of any apparent absorption at $\sim
5017$\AA~ suggests that the \civ~ absorption column is significantly
lower (and is consistent with no \civ~absorption at all) in this shell
as compared to that associated with absorber \#2, and we assume that all
of the \civ~ absorption can be accounted for by associating it with
\Lya~ absorber \#2. This implies a metallicity gradient in the
environment of this radio source. Overzier et al. (2001) recently
found strong evidence for a metallicity gradient within the gaseous
halo of the $z = 2.49$ radio galaxy MRC2104-242. Overzier et
al. (2001) attributed this metallicity gradient to a scenario in which
the gas is associated with a massive cooling flow or originates from a
merging event.  This scenario can also be extended to the absorption
gas, and we return to this point in section~\ref{sec:discussion}.
However, one should keep in mind that the measured $N_{\rm CIV}$ is an
upper limit, due to possible extra absorption from the other
absorption shells.

\subsection{The metallicity of the absorbing shell}

In this section we investigate whether the absorption- and
emission-line gas can exist co-spatially or whether they are
distributed differently based on ionisation models of gaseous slabs
with the column densities we observe.  We use the method of B00 to
investigate the metallicity of the absorbing shell \#2 which is the
only one clearly seen in absorption in
\civ. We use the observable quantity \gam\ as described by B00 to
describe the various ratios between the emission- and the
absorption-line properties of both Ly$\alpha$ and \civ. $\Gamma$  is defined as
\begin{equation}
\gam = {I_{CIV}\over I_{Ly\alpha}} {N_{HI} \over N_{CIV}}  , 
\end{equation}
where ${I_{CIV}/I_{Ly\alpha}}$ is the ratio of the emission-line
fluxes and $\nhi/\nciv$ is the ratio of the
measured absorption columns. These four quantities carry information
on the three ionisation species H$^0$, H$^+$ and C$^{+3}$.

\subsubsection{Co-spatial absorption and emission line gas}
Using the current spectra and the Gaussian fit to the emission lines,
we obtain ${I_{CIV}/I_{Ly\alpha}} = 0.12 \pm 0.10${\footnote{The error quoted is the $1\sigma$ uncertainty derived from the symmetric and asymmetric Gaussian fits to the emission-lines of both Ly$\alpha$ and \civ.}. Adopting the column of
\Lya~absorber \#2 in Table~\ref{tab:Lya0200main} and that of CIV in Table~\ref{tab:civ0200}, we obtain $\nhi/\nciv
= 1.09 \pm 0.04$, hence $\gam \approx 0.13 \pm 0.12$. This is a factor of $\sim 5 \times
10^4$ smaller than the value found by B00 for 0943--242 of $\gam =
5400$.  B00 used the high
\gam\ value in 0943--242 as evidence that the absorbing gas was physically
distinct from the emission line regions and was of much lower density
and metallicity. Calculations with the \map\ code (Ferruit et
al. 1997) indicate that our extremely small value of \gam\ effectively
rules out that the same solar metallicities characterize both the
emission- and the absorption-line regions. In fact, reproducing a
\gam\ of 0.13 requires a metallicity of $\sim 10~Z_{\odot}$ if we insist
that the absorption and emission gas components have the same
metallicities. Therefore, it is theoretically possible to reproduce
\gam\ with a strong overabundance of metals, which may suggest that
the emission gas and absorption gas might be more directly
inter-related in 0200+015 than in 0943--242. On the other hand, the
fact that the emission has such a large FWHM as compared to the $b$
values of the absorption profiles and the near unity covering factor
of the absorption gas, would still favour the simple picture in which
the absorption gas is more quiescent, possibly situated in
quasi-uniform shells which are either expanding or infalling.

\subsubsection{An absorbing shell surrounding the emission-line region}
We now consider the absorption gas in isolation and attempt to explain
the process by which the \civ\ column can be approximately equal to that of
\hi\ in the case of absorber \#2.  At a given metallicity, there exists
a physical upper limit to the $\nciv/\nhi$ column ratio which
photoionisation models cannot exceed. This is shown in
Fig.~\ref{fig:matt} which show photoionisation calculations by an
ionizing continuum of photon index $\alpha = -1$ ($F_{\nu} \propto \nu^{\alpha}$) as a function of the
ionisation parameter\footnote{$U = \frac{Q}{4\pi r^{2}~c~n}$, where $Q = \int \frac{L_{\nu}}{h\nu}~d\nu$, $r$ is the distance to the cloud from the ionising source and $n$ is the particle density.} \up. The geometry considered was that of an
optically thin slab with all models having the same thickness in \nhi\
of $10^{14.73}$\,\cms. Inspection of Fig.~\ref{fig:matt} shows
that the solar metallicity curve cannot reach the observed value
of 0.9 while a slab of $10~Z_{\odot}$ can again reproduce the target
ratio. If the ionisation process, however, was dominated by
collisional ionisation, then one could not rule out solar
metallicity.
This is shown by the dashed-line in Fig.~\ref{fig:matt}
which represents a slab which is initially photoionised with a low value of
\up\ but where the temperature is imposed. The
temperatures are increasing (starting from the equilibrium value of
13000\,K) in steps of 0.1\dex\ and collisional ionisation
progressively takes over. The open squares identify the first eight
models, up to $8.2 \times 10^{4}\,$K. To show that the column ratio
actually decreases at much higher temperatures, we used filled
triangles to denote the models in which the temperature starts from $1.03 \times 10^{5}\,$K and increases in steps of 0.1~dex.  We have verified that, as expected, the
peak reached in the column ratio is not higher for a slab with higher
ionisation parameter (the peak is actually lower for $U \ge 0.01$).

We infer from the above calculations that a solar metallicity would therefore require an unusual fine tuning in order to give a temperature which lies in the range of $\sim 0.7$--$1.4 \times 10^{5}\,$K.
This is not a very probable state for any
astrophysical plasma since this is exactly the same temperature range
across which the cooling curve is known to peak (e.g. Fig.~8 in Ferruit 
et al. 1997). For the gas to
be maintained in this temperature range would require an additional (unknown) heating
mechanism.  Cooling shocks can in principle reproduce the
observed \civ\ columns but the models so far computed indicate huge
\nhi\ columns $\sim 10^{19}\,\cms$ (see Table 2D in Dopita \&
Sutherland 1996), a value which is many orders of magnitude beyond the
observed value for absorber \#2 of $10^{14.73}$\,\cms. For these
reasons, it appears to us more plausible to invoke metal-rich
absorbing gas.

Therefore, although the value of $\Gamma$ in 0200+015 (unlike 0943-242
cf. B00) can be reconciled with a co-spatial distribution of
absorption- and emission-line gas, this is not our preferred scenario
given the additional information, i.e. the fact that the emission has
such a large FWHM as compared to the $b$ values of the absorption
profiles and the near unity covering factor of the absorption gas.
Therefore, the original conclusions of B00 still hold in which the absorption gas resides in a shell-like distribution, but we are now
in a situation where the metallicity and the density of the
absorption shells varies from shell-to-shell in the same source and
also from source-to-source.

\begin{figure}
\resizebox{\hsize}{!}{\includegraphics{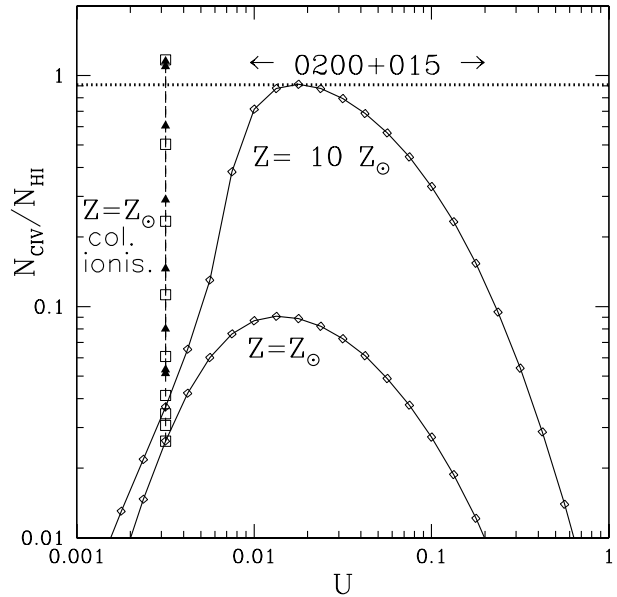}}
\caption{The column ratios $\nciv/\nhi$ in 
photoionisation models as a function of the ionisation parameter $U$
(represented by open diamonds joined by a solid line).  The lower and
upper curves correspond to gas with solar and 10 times solar
metallicity, respectively. The dashed-line corresponds to models whose
temperature has been increased in steps starting from the
equilibrium value of 13000\,K (open squares for $T \le 82000\,$K,
filled triangles at higher $T$). The $\nciv/\nhi$ column ratio
actually decreases when the temperature exceeds $10^{5}\,$K. All
models correspond to slabs for which the \nhi\ column is $10^{14.73}$\,\cms\ as in 0200+015 (absorber \#2 in Table 2). The dotted
horizontal line represents the observed value of $\nciv/\nhi$ ratio in
0200+015.}
\label{fig:matt}
\end{figure}

\section{DISCUSSION}\label{sec:discussion}

\subsection{Origin of the absorbing shells}
In this section we speculate on the origin of the large absorption
shells around HzRGs. 

\subsubsection{Shells propagated by the passage of radio jets}
The evidence presented in this paper suggests that the absorbing
shells surround the whole of the radio source and probably have a
roughly spherical distribution. If this is the case then it is extremely
unlikely that the shells are propagated by the passage of the radio
emission for which we would expect the shells to be disrupted at radii
less than the extent of the radio emission. The fact that we see the
absorption systems in the majority of the smaller radio galaxies (projected linear size $D < 50$~kpc) but not in the larger sources (e.g. van Ojik et al. 1997) makes this scenario very unlikely. 
We also observe the absorption systems along the line-of-sight to the nucleus. Given that radio galaxies are orientated along the plane of the sky according to orientation based unified schemes, then this also rules out the
scenario of a radio jet propagated shell.

\subsubsection{Starburst driven superwinds}

One explanation is that the shells originate
at the centre of the galaxy and are ejected by superwinds, like those
seen in Lyman Break galaxies (e.g. Steidel 2001). These winds
will propagate roughly spherically, provide an explanation of the
enriched nature of the shells and have no problem in explaining the
number of shells and the diverse metallicities if the starburst driven
winds are episodic.

Evidence is mounting that powerful radio galaxies are formed at
an early epoch and the triggering of the AGN activity is synchronized
with a major episode of star formation (e.g. Archibald et al. 2001;
Willott et al. 2002; Baker et al. 2002). Therefore, it seems likely
that starburst driven superwinds may be an important mechanism for
ejecting gas out of the central regions and the high dust concentrations inferred from submillimetre observations imply that starburst may easily be obscured at UV and optical wavelengths.

Martin (1999) and Heckman et al. (2000) estimated the outflow speeds
of superwinds for galaxies of varying mass. They show that the outflow
speed is largely independent of galaxy mass, implying that the
outflows preferentially escape less massive halos.  High-redshift
radio galaxies seem only to be found in the most massive elliptical
galaxies at any redshift (e.g. Jarvis et al. 2001b), hence presumably
the deepest potential wells. If this is the case then unlike the case
for $L^{\star}$ and sub-$L^{\star}$ galaxies the winds may not possess
the required velocity to fully escape the deep potential well and subsequently enrich
the intergalactic medium beyond the confines of the HzRG dark
halo. Instead this enriched material resides at large radii, away from
the central cusp, but still within the potential well of the
galaxy. 

In the event of episodic superwinds, these shells would eventually
collide at large radii, which could both deplete or increase the
metallicity of the most distant shell depending on the composition of
the wind.  If this is the case then these shells will eventually fall
back towards the centre of the shell in the absence of a repulsive
force.  Indeed, with reference to Tables~\ref{tab:Lya0943} and
\ref{tab:Lya0200main}, there is a slight tendency for the
\Lya~absorbers in 0200+015 to have lower neutral column densities and
$b$-parameters as the absorption redshift ($z_{\rm{abs}}$)
increases. If the difference between $z_{\rm{abs}}$ and the redshift
of the emission line peak reflects outflow of the shell relative to
the radio galaxy, this trend could reflect evolution in the ionisation
state and internal structure of the shells as they are ejected from
the radio galaxy. 

This is in line with an initial major burst of star
formation in which much of the gas is expelled, subsequently followed
by smaller bursts in which the lower velocity, lower-density shells
are expelled. At some stage in the evolution these shells would fall
back under gravity and the shells would merge, but this process is
probably very rare because the radio jet has already disrupted the
outermost shells which are now seen in emission.
However, it is difficult to reconcile the low metallicity shells in
0943-242 with a starburst scenario, as we would expect much higher
enrichment if the shells originate in a starburst driven
superwind. Although this is not the case for the high metallicity
shell in 0200+015.

\subsubsection{Relics from a gas-rich merger}
Metal enriched shells of low-density gas could be a relic from
a gas-rich merger in which there was an episode of star-formation,
presumably induced by the merger.
This scenario has been used to explain the vast \HI~ shells in low-redshift elliptical galaxies (e.g. Sahu et al. 1996; Morganti et al. 1997).

A number of numerical
investigations predict that the outer regions of a galactic halo
should be less metal rich than the central regions (e.g. Bekki
1998). This is in line with the case of 0943-242 but in opposition to the case of 0200+015 in which the main
absorber is highly enriched and is comparable in metallicity to that
expected in the narrow-emission line region towards the centre.

An argument against the merger relic scenario is that the efficiency
of mergers depends critically on the relative velocities, with low
velocity mergers being the most efficient.  This means that these
efficient mergers would produce very low velocity dispersions in the
absorbing gas. We might also expect
the low density halo resulting from a gas-rich merger to have a more
filamentary structure with a covering factor less than unity, which although reconcilable with our observations of a unity covering factor along
our line-of-sight, is not the most plausible explanation. Therefore although a gas-rich merger remains a
candidate for producing the extended low-metallicity haloes it has
difficulties in producing the enriched shell we see in 0200+015 and
probably needs a very specific merging scenario.

\subsubsection{Gas-rich clusters and cooling flows}

It is now well established that many powerful radio galaxies reside in
significant galaxies overdensities at all cosmic epochs (e.g. Hill \&
Lilly 1991; McLure \& Dunlop 2001; Best 2001; Pentericci et
al. 2000). Indeed, the discovery of $\sim 20$ Ly$\alpha$ emitters around a $z
= 4.1$ radio galaxy (Venemans et al. 2002) suggests that powerful
radio galaxies are ideal beacons for tracing proto-clusters in the
early Universe. If this is ubiquitous for all powerful radio galaxies
then we might expect large reservoirs of gas around the radio galaxy.

Within a cluster, cooling flows provide a mechanism for depositing substantial 
amounts of cool gas and dust around the central galaxy (e.g. Edge 2001). 
The observational consequences of a radio source being triggered at the centre 
of such a cooling flow cluster at high redshift were discussed by Bremer et al. (1997). One of their clearest predictions concerns the 
presence of \HI~ absorption in the \Lya~emission in systems seen as radio 
galaxies (as opposed to radio-loud quasars): to see absorption from an 
ensemble of clouds requires a covering fraction of at least 50 per cent; 
little absorption is expected from clouds at radii of 10--100\kpc, as their 
covering fraction is low, but those between 10--20~\kpc~provide a covering 
fraction of order unity and absorbing \HI~ columns of $10^{18}-10^{21}$\psqcm, as the surrounding gas away from the ionising beam should be similar to that of the interstellar medium before the nucleus became active.. 
This would account for the absorption in 0943-242 where the radio source has 
a projected linear size $\sim 26$\kpc; in 0200+015, however, the radio source 
is larger ($\sim 43$\kpc) and has already engulfed these high-covering factor 
clouds, so the absorbers in this object may have been disrupted.

However, the classical view of a cooling flow, from low-redshift
studies is almost certainly not applicable at these high
redshifts. The main reason for this is that the environmental gas
may not have had time to both settle towards hydrostatic equilibrium
and also to be heated up to X-ray temperatures. In the low-redshift
Universe cooling flows also have an inclination to produce filaments
when the gas cools (e.g. Fabian et al. 1981), which would result in a much reduced
covering factor ($\ll 1$; see e.g. Conselice, Gallagher \& Wyse 2001). Therefore, a possibly more accurate view is
one in which the pristine gas is collapsing and cooling in a more
globally ordered fashion, i.e. as a spherical collapse. In this case
the resulting gaseous distribution would be different from that in a
low-redshift cooling flow and have the observed high covering factors.

\subsection{A scenario: deposition of pristine gas and triggered star formation}
An age based scenario, which assumes that both the AGN
activity and a major episode of star formation are synchronized,
probably due to the collapse and cooling of (proto-)cluster gas (or a gas-rich merger), may
explain the differences in the absorption shells of 0200+015 and
0943-242. The projected linear size $D$, of the radio emission may be used
as an age indicator for the radio source (e.g. Blundell, Rawlings
\& Willott 1999; Baker et al. 2002). In this case 0200+015 ($D \sim
43$~kpc) is older than 0943--242 ($D \sim 26$~kpc). If
this is the case then there is much more time for the absorption
shells to have evolved in 0200+015. An indication of this comes from
the interaction of the radio emitting plasma in the extended component
in 0200+015, which looks to be disrupting the absorption shells and
subsequently producing additional emission. 

If starburst or quasar driven winds precede the passage of the radio
activity then a pristine halo at large radii will be enriched by the
winds before the radio plasma propagates through the extended
regions. Indeed, in an investigation of the metallicity of the
broad-line region in quasars, Hamann et al. (2002) compared the UV
emission line ratios in a sample of radio-quiet quasars with
theoretical models. They found the broad-line region gas was most
likely to be of high metallicity ($Z > Z_{\odot}$) and was a
consequence of rapid star formation triggered at or before the time of
the quasar activity. Ejection of the gas via a quasar or starburst
driven outflow would obviously enrich the surrounding halo and thus
offer an explanation of the highly enriched nature of absorber \#2 in
0200+015.

In the case of 0943--242 any
starburst driven wind associated with the activation of the nucleus
may not have had time to enrich the outer halo. Thus the
explanation of Binette et al. (2000) still holds for this source, in
which the shell has a very low-metallicity. 

This would also explain the variety of metallicities observed in these
shells, as some of the shells will have been enriched, while the
winds have not yet reached the others. This naturally fits in with the
absorption lines in 0200+015, in which the the blueshifted (absorber
\#2) shell has a high metallicity, whereas absorbers \#3 and \#4 lie redward
of the Ly$\alpha$ emission line and are probably infalling components from
the (proto-)cluster environment, in which there has been little or no star
formation to enrich the shells. This is also in line with the work of Overzier et al. (2001) in which the metallicity gradient in the emission-line gas in the gaseous halo surrounding MRC2104-242 may also arise from the same mechanism.

The fate of the shells is uncertain, it
is not implausible that shredding by the passage of the radio jet (e.g. Bremer et al. 1997) will work to condense
the shells into distinct clouds that may resemble the knots of
emission-line gas observed in many larger radio sources. This
naturally fits in with correlation between the extent of the
emission-line gas and the extent of the radio emission (e.g. Jarvis et
al. 2001a), with the radio emitting plasma propagating through the
low-density gas which consequently increases the density of this
gas resulting in detectable line emission.

\subsection{The link to quasars}

Using a complete sample of radio-loud quasars from the Molonglo Quasar
Catalogue (Kapahi et al. 1998), Baker et al. (2002) used both the Hubble
Space Telescope with STIS and ground based telescopes to investigate
the associated \civ~ absorption in two redshift intervals.  They found
that the absorbing systems in these objects have similar velocities
widths and offsets to the two radio galaxies investigated in this
paper, the gas is mixed with a large amount of dust and that the gas
and dust dissipate over the lifetime of the radio source.  Thus, the
main conclusions from this study fit in easily with our results, in
line with Unified schemes.

The advantage that radio galaxies hold over quasars is that, according
to unified schemes, the radio galaxies are orientated along the plane
of the sky and so the deprojection to the true linear size of the
radio source is trivial. Therefore, age effects may become more
apparent in a large sample of radio galaxies with higher
signal-to-noise observations. This should easily be achievable with
current instrumentation as we have shown that spectra of resolution
of $\sim 2$\AA~ at these redshifts is sufficient to probe the
absorption shells to the required detail allowing the integration time
to be significantly reduced. Combined with a similar sample of
radio-loud quasars, in which the reddening properties may be explored,
we should be able to track the evolution of both the absorption- and
emission-line gas and the intrinsic dust properties in all radio-loud AGN.

\subsection{Further observations}
Obviously further observations of the sources studied in the paper
will enhance our understanding of the origin and fate of the gas
surrounding HzRGs. Indeed, deep spectroscopy around other resonant
lines may provide critical information regarding the state of the
absorbing shells. The most obvious emission line to use for this study
would be the Mg{\footnotesize{II}} $\lambda\lambda$2795,2802
doublet. The absorption equivalent width ratio of
\civ/Mg{\footnotesize{II}} is frequently used to classify quasar
absorbers into high or low excitation (e.g. Bergeron et al. 1994) and will also provide further information regarding the metallicity of the gas.

0200+015 is also at a redshift in which the H$\alpha$ emission line is
redshifted into the near-infrared $K-$band transmission window. As
H$\alpha$ is not a resonant line the true velocity structure of the
Ly$\alpha$ line should be replicated in H$\alpha$, thus allowing a
detailed study of the actual kinematics of the emission-line gas. The
Ly$\alpha$/H$\alpha$ ratio is also sensitive to reddening by dust, and
if the Ly$\alpha$ photons escape from the clouds via resonant
scattering from H{\footnotesize{I}}, the Ly$\alpha$ line should have
significantly broader wings than that of H$\alpha$.

Finally, spectroscopic observations around various resonant line of a
large sample of high-redshift radio galaxies spanning projected linear
sizes from $D \sim $1-2~kpc up to the classical doubles with $D~ \gtsim~
100$~kpc are warranted. These will shed further light on the
distribution, density and metallicity of gas in the environments of
the most massive galaxies at high redshift, and provide clues as to
the origin and fate of these large gaseous halos, and may consequently
provide important clues to galaxy formation in general and the
important role that feedback may play.

\section{Conclusions}\label{sec:conc}
We have obtained high-resolution spectra of two high-redshift
radio galaxies to determine the structure of the absorption gas which
appears to be ubiquitous in small ($D < 50$~kpc) radio sources.
The main conclusions of this study can be summarized as follows:
\begin{itemize}

\item  The higher resolution observations do not uncover significant structure in the absorbers of both Ly$\alpha$ and C{\footnotesize{IV}} in 0943-242, however the $N_{\rm HI} \sim 10^{19}$~cm$^{-2}$ absorber in 0200+015 split into two absorber of $N_{\rm HI} \sim 4 \times 10^{14}$~cm$^{-2}$. The absorption gas present in our two chosen high-redshift radio sources
is most likely distributed in shell-like structures which encompass
the radio source.

\item The metallicity of these absorption shells can vary between sources
and also between the shells in the same sources, implying that the
shells are enriched by a secondary process.

\item We speculate that the most plausible origin of the
absorbing shells is two fold. The pristine shells are the relics from a gas-rich merger or (proto-)cluster gas which has cooled and collapsed towards the centre of the dark matter halo. Whereas the enrichment comes at a
later stage when the AGN is triggered simultaneously with a major
episode of star formation, presumably due to a large inflow of gas into the nucleus. This major burst of star formation
initiates starburst driven superwinds which drive the gas
toward the outer halo of the galaxy and may explain the very high metallicity of the absorption shells in 0200+015.

\item The passage of the radio emitting plasma disrupts and fragments the
absorption shells in the larger, and thus older radio
sources. Evidence of this is seen in 0200+015, where the absorption screen along our line-of-sight to the radio lobe appears to be
more fragmented than the absorbing screen between us and the nuclear
emission.

\item We have found direct evidence that the narrow-emission lines in
high-redshift radio sources may be outflowing from the central
regions. This is in agreement with previous studies of the
narrow-emission line dynamics in nearby Seyfert galaxies and is in
line with orientation based Unified Schemes in which the outflow from
the far side of the source is obscured by dust.

\end{itemize}

\section*{ACKNOWLEDGMENTS}
We thank the staff at the Paranal Observatory for their excellent
support and Sandro D'Odorico for help with the preparation for the UVES observations. RJW thanks Sara Ellison and Sabine Mengel for their
assistance with the data reduction. We thank Jaron Kurk for the use of
his IDL rountine for Voigt profile fitting. MJJ acknowledges the
support of the European Community Research and Training Nework "The
Physics of the Intergalactic Medium". RJW acknowledges support from an EU Marie Curue Fellowship. LB acknowledges support from the CONACyT grant 32139-E. The data presented in this paper
were based on observations performed at the European Southern
Observatory, Chile (Programme ID: 68.B-0086(A)).

{}

\end{document}